\documentclass{aastex}

\slugcomment{Accepted by Ap.J. Letters}

\shorttitle{ADAFs don't match kepler disk}
\shortauthors{Molteni et al.}

\begin{document}


\title{WHY CANONICAL DISKS CANNOT PRODUCE ADVECTION DOMINATED FLOWS}


\author{D. Molteni\altaffilmark{1} , G. Gerardi\altaffilmark{1} and M. A. Valenza\altaffilmark{1}}
\affil{Department of Physics and Related Technologies, University of Palermo, Italy, 90100}

\email {molteni@gifco.fisica.unipa.it}

\altaffiltext{1}{Department of Physics and Related Technologies, University of Palermo, Italy, 90100}


\begin{abstract}
Using simple arguments we show that the canonical
thin keplerian accretion disks cannot smoothly match any plain advection
dominated flow (ADAF) model. By 'plain' ADAF model we mean the ones with zero cooling.
The existence of sonic points in exact solutions is critical and imposes
constraints that cannot be surpassed adopting 'reasonable' physical conditions
at the hypothetical match point. Only the occurrence of new critical physical phenomena may produce a transition. We propose that exact advection models are a
class of solutions which don't necessarily involve the standard thin cool disks and suggest a different scenario in which good ADAF solutions could eventually occur.
\end{abstract}


\keywords{: accretion disks---black hole physics---hydrodynamics}


\section{Introduction}

Recently, the need to explain black hole
candidate spectra and the low luminosity of some galactic nuclei, supposed
to contain a black hole, prompted more attention to advective accretion
disk models \citep{narai95}  The basic idea is that the gravitational
energy content doesn't go into local radiation emission as in standard thin
and cool keplerian disks (let us call SSD) \citep {ss73}, but goes into
heating and into radial motion. Therefore a new class of disks has been
proposed:\ ADAF (Advection Dominated Accretion Disks). These disk models
take into account the advection and pressure terms in the
fluid-dynamics equations. Indeed, exact solutions of accretion flows
(including viscosity, but only for isothermal cases) have been found 
years ago by S. Chakrabarti (1990). Literature has been enriched by many
kinds of such advective disks with different physical properties, see
Bisnovatyi-Kogan (1999) for an extensive review.
We will examine the very simple ADAF models
initially proposed (Narayan and Yi 1995; Narayan, Kato and Honma, 1997;
Igumeshev, Abramowicz and Novikov, 1998;
Dullemond and Turolla,1998; Lu, Gu and Yuan, 1999),
i.e. the ones without any cooling. As stated by
the authors, although this model may appear very crude, it contains the basic
ingredients: the role of pressure gradient and of the advection term.
Analytical solutions have been obtained only for the self similar case
\citep {narai95}, but this approximation is not tenable for the case of
BH accretion, since all standard ADAF solutions rely on the conditions at the 
sonic point, which is close to the BH.

Other solutions have been typically obtained with numerical integration of
the steady state equations. The most used technique is the relaxation method
with boundary conditions given by 'reasonable' physical approximations. In
all works it is under laying the meaningful but 'a priori' idea that 
necessarily the ADAF must spring out of a canonical disk. So all the numerical
procedures compel the solutions to fit the standard kepler disk. It is
worthwhile to add that it is well known that the relaxations methods can
relax onto unphysical solutions.

A simple energy consideration gives a cutting argument against the possible smooth
convergence of ADAF to SSD disks. 
The fact that self similar solution by Narayan and Yi cannot be automatically 
connected to cold thin disks has already been indicated  by Kato and Nakamura (1998). 
Here we point out that even the more sophisticated (not self similar) ADAF 
cannot have an asymptotic approach to the SSD models.
We conclude that,  possibly, standard 'plain' ADAF models are 'numerically forced'
solutions
and the claimed asymptotic approach to a keplerian disk is not
demonstrated. 

\section{Basic equations}
In the 'plain' ADAF model it is
assumed that it doesn't contain any diffusive term and the
cooling processes are irrelevant, as occurring on timescales
much longer than the advection fall time. No heat conduction is considered
as in Narayan and Yi (1995), Narayan, Kato and Honma (1997), Dullemond and
Turolla (1998). Then the basic
equations are the ones given by classical viscous fluid dynamics. We adopt
newtonian physics and the black hole force on the fluid can be derived from
the appropriate Paczy{\'n}ski and Wiita potential (1980). This is also a
standard procedure in many papers. On the other hand the classical approach
makes more clear the basic physical point without inessential complications
due to an exact relativistic treatment.

Therefore an exact disk solution has to obey the following equations:

Mass conservation 
\[
\stackrel{.}{M}=r\Sigma v_r=const
\]

Radial momentum equation 
\[
v_r\frac{dv_r}{dr}=-\frac 1\Sigma \frac{d\Pi }{dr}-G\frac{M_{*}}{\left(
r-r_g\right) ^2}+\frac{\lambda ^2}{r^3}
\]
where $\Sigma $ and $\Pi $ are the vertically integrated density and
pressure,

Stress definition 
\[
\tau _{r\phi }=\eta r\frac{\partial \Omega }{\partial r}
\]
where $\eta =\alpha \rho aZ_{disk}$      with  $a=\sqrt{\gamma \frac
P\rho }$ is the sound speed,

Angular momentum equation

\[
\frac{\rho v_r}r\frac{d\lambda }{dr}=\frac 1{r^2}\frac \partial {\partial
r}\left( r^2\tau _{r\phi }\right) 
\]

that leads to 
\[
\stackrel{.}{M}\lambda =4\pi r^2\tau _{r\phi }+\stackrel{.}{J}
\]

Vertical equilibrium gives the disk thickness
\[
Z_{disk}=\frac a { \Omega_k }
\]

Energy equation 
\[
\frac{d\varepsilon }{dt}=v_r\frac{d\varepsilon }{dr}=-\frac P\rho \frac 1r%
\frac{d\left( rv_r\right) }{dr}+\frac \Phi \rho 
\]
with $\varepsilon $ thermal energy per unit mass, $\Phi $ the dissipation
function given by : $\Phi =\frac{\left( \tau _{r\phi }\right) ^2}\eta $ .

Equation of state 
\[
P=\left( \gamma -1\right) \rho \varepsilon 
\]

 $\Omega_k $ is the keplerian angular velocity, all other terms have their standard 
meaning accepted in the referenced papers.

\section{Boundary conditions }
With the help of the energy equation,
that will be discussed in the next section, our problem is reduced to solve a set
of 3 algebraic equations (vertical equilibrium, mass, energy) and two differential
equations (radial momentum and angular momentum). 
The existence of sonic points, requiring the continuity of the variables and of the 
speed derivatives, links these boundary conditions. In particular, once the sonic point is 
determined, then also the derivative of the radial speed at that point is fixed.
The derivation of the analytical formula of $\frac{\partial V_r }{\partial r}$ at the sonic
point is algebraically complicated but straighforward.
The usual assumption $\frac{\partial \lambda }{\partial r}=0$ at the sonic point
is reasonable only for sonic point close to the BH and for low viscosities, and it is
not true in general. Due to the different approaches to the numerical solutions
(relaxation from fixed inner and outer points, standard Runge-Kutta integration from sonic point) 
there is no consensus on the exactness of the adopted boundary conditions.
Our solutions are computed integrating the equations starting from the sonic point 
with the appropriate analytical derivative values calculated by the 
De L'Hopital rule. 
Artemova et al. (2001) using a similar treatment of the sonic point
conditions have obtained subkeplerian flows starting from canonical disks.
However their results
don't disprove our conclusions, since they refer to disks
obtained taking into account
radiation transfer and using different viscosity prescriptions.
Apart any further consideration on boundary conditions, we show that some crucial
conditions are imposed by the energy equation.




\section{ The energy argument }

An interesting point, apparently not exploited in ADAF context, although
appearing in many papers ( Hoshi 1984, Honma 1996) is the fact that it
is possible to have a constant energy property, reformulating the energy
equation in the floowing way.

\[
div\left[ \rho {\bf v}\left( \frac 1 2{\bf v}^2+\Psi \left( {\bf r}\right)
+h\right) -{\bf v}:{\bf T } \right] =0
\]

$h=\frac P\rho +\varepsilon $ is the enthalpy, {\bf T } is the stress tensor. 

This expression, after insertion of the stress definition, integrating over Z, and using 
the mass conservation equation,
{\em in general} gives the
following relationship, even in the differential form of the viscosity tensor: 

\[
\frac 1 2 v_r^2-G\frac{M_{*}}{\left( r-r_g\right) }+\frac{a^2}{\left( \gamma
-1\right) }-\frac{\lambda ^2}{2r^2}+\frac{\lambda \lambda _e}{r^2}=\frac{%
const}{\stackrel{\cdot }{M}}
\]

$\lambda _e$ is the angular momentum at the inner edge of the disk, usually set equal to 
the value corresponding to the last stable orbit around the Black Hole.

The energy equation is valid over the whole ADAF
solution up to the eventual
crossing point with the keplerian disk. A smooth transition between ADAF and
SSD\ disk requires
a continuous smooth limit of the ADAF quantities to the SSD quantities:
temperature, radial speed, density, etc.;

What is crucial in this formula is that, at the crossing point of the
solutions, where the angular momentum is keplerian, the following relation is
obviously valid 

$$
\frac{\lambda _k^2}{2r^2}\gg \frac{\lambda _k\lambda _e}{r^2}
$$

since it is reasonable and common to choose $\lambda_e \ll \lambda _k$.
Now, if the ADAF solution has to smoothly match a standard keplerian disk at
a radial distance far from the BH $\left( r\gg r_g\right) $ , we can
substitute $\lambda $ with $\lambda _k$; another
reasonable and
usual assumption is that at the match point the flow be subsonic \citep{ig98}.

Therefore at the crossing point $r_{+}$
we must have : 

\[
\frac{a^2}{\left( \gamma -1\right) }=K+\frac{GM_{*}}{r_{+}}
+\frac{\lambda _k^2}{%
2{r_{+}}
^2}=K+\frac 32\frac{GM_{*}}{r_{+}}
\]

Now, requiring that the energy constant be of the same (or of the same order) of
the energy per unit mass of a standard keplerian disk, i.e. $K=-\frac 1 2%
\frac{GM_{*}}r$ , we have :

\[
\frac{a^2}{\left( \gamma -1\right) }=\frac{GM_{*}}{r_{+}}
\]

$\frac{GM_{*}}{r_{+}}$
corresponds to the virial temperature , much larger than the standard temperature
of a canonical thin keplerian disk. 
We have a strong inconsistency: the
ADAF temperature, at the matching point, is  always much larger than the
canonical keplerian disk temperature (Frank, King and Raine 1992).

Indeed  Lu et al. (1999) use our same numerical approach (i.e. starting from sonic
point) and apparently they find cases with a smooth convergence from ADAF to 
SSD; but it is worth nothing that their parameter space, corresponding to 
Narayan ADAF model, has zero measure. 
Those configurations lie on a line in their parameter space $R_{s}$ , $\lambda _e$.
Therefore they can be obtained
only for a single $\lambda _e$ if the sonic point is given.

\section{Solutions}
We performed several integrations using the direct
method, starting from the sonic point inwards and outwards.
Figure 1. shows the result of the Runge-Kutta integration of our equations.
It is plotted the angular
momentum per unit mass versus the radial distance for $\alpha = $%
0.005, 0.01, 0.02. As the viscosity increases the angular momentum crosses
the keplerian angular momentum at distances closer to the BH.
Figure 2. shows the sound speed and the Mach number for the same $\alpha$ values.
The sound speed at the crossing point doesn't go to extremely low values. The values
we obtain are exactly the ones predicted by the energy equation and
go to a finite limit at a finite radius located beyond the keplerian crossing point.
The angular momentum of  ADAF solutions crosses the kepler angular momentum
value in a non smooth way. 
The exact solution continues beyond that point and stops at a larger distance with disk conditions (temperature, density, radial speed) not corresponding to the canonical keplerian disk. 
Our correct ADAF solutions are 'per se' standing solutions.
This kind of solutions is a particular case of the more general solutions
drawn by Chakrabarti and Titarchuck (1995) and Chakrabarti (1996), who explored a more general
case including an 'ad hoc' cooling treatment.
We add a further comment on the shock or no shock solutions, since many authors consider
the Chakrabarti solutions untenable as the shock he finds had not been confirmed by their 
calculations.
As has been pointed out 
in early works (Liang and Thompson, 1980) there are many sonic points in the flow solutions. 
The flow on a black hole has no rigid surface
and shock occurs only in solutions passing through the outer sonic point as clearly
demonstrated by Chakrabarti (1990) and Chakrabarti and Molteni (1993).
All the ADAF solutions we are discussing pass through the inner sonic point, so it is
quite obvious that no shock can occur in these branches: the flow is supersonic only
after the sonic point which is close to the black hole horizon.
Lanzafame et al. (1998) tested these solutions with a time dependent code
and found that these
solutions are stable, while the standard ADAF\ solutions have been
demonstrated unstable by numerical simulations (Igumenshchev and Abramovicz, 1996).

\section{Conclusions}

We conclude that plain ADAF\ solutions that don't
respect the full set of sonic point conditions, including derivatives at sonic point, are 
incorrect.
However, since ADAF models seem to offer explanation of relevant phenomena, to account for their
existence, and in particular for the formation of a hot comptonizing corona,
we propose that a subkeplerian thin flow is existing by itself, superposed to
the standard keplerian disk. In accretion occurring in Active Galactic Nuclei this subkeplerian 
flows  could easily come from the star cluster, that is known to have a very slow average rotation \citep{ho99}. 
In low mass binary system this subkepler flow 
could come from a wind induced by the high energy radiation impinging on the normal 
star surface, or from gas recirculating in the binary potential well \citep{bisi98}. 
We are also testing the hypothesis that a physically motivated change in Z of the viscosity
prescription can lead to differentially rotating flows.





\acknowledgments

\clearpage



\figcaption[angmom.eps]{Angular momentum per unit mass versus radial distance for $ \alpha =.005,.01,.02$. \label{fig1}}

\figcaption[masosp.eps]{Mach number and sound speed versus radial distance for $ \alpha =.005,.01,.02$. \label{fig2}}



\begin{thebibliography}{}
\bibitem [Artemova et al.(2001)]{art01} Artemova I. V., Bisnovatyi-Kogan G. S., Igumenshchev I. V., Novikov I. D., 2001, in press \apj, (astro-ph/0003058)
\bibitem [Bisnovatyi-Kogan (1999)]{bi99} Bisnovatyi-Kogan G., 1999, in  Observational Evidence for Black Holes in The Universe, S.K.Chakrabarti Editor (Dordrecht Kluwer Academic Press),1.
\bibitem[Bisikalo et al.(1998)]{bisi98} Bisikalo A.A., Boyarchuk V.M., Chechetkin, Kuznetsov O.A., Molteni D., 1998, \mnras, 300, 39
\bibitem [Chakrabarti (1990)]{ch90} Chakrabarti S.K., 1990, Theory of Transonic flows, Singapore, World Sci.
\bibitem[Chakrabarti(1996)]{ch96} Chakrabarti S.K., 1996, \apj , 464, 664
\bibitem[Chakrabarti and Molteni(1993)]{chm93} Chakrabarti S.K., Molteni D., 1993, \apj, 417, 671
\bibitem[Chakrabarti and TitarchuK (1995)]{ch95} Chakrabarti S.K., Titarchuk L., 1995, \apj, 455, 623
\bibitem[Dullemond and Turolla(1998)]{du98} Dullemond C.P., Turolla R., 1998, \apj , 503, 361
\bibitem [Frank et al.(1992)]{fr92} Frank J., King A., Raine A., 1992, Cambridge University Press
\bibitem [Ho(1999)]{ho99} Ho L.C., 1999, in  Observational Evidence for Black Holes in The 
Universe, S.K.Chakrabarti Editor (Dordrecht Kluwer Academic Press), 153
\bibitem[Honma(1996)]{hon96} Honma F., 1996, \pasj, 48, 77
\bibitem[Hoshi(1984)]{hos84} Hoshi R., 1984, \pasj, 36, 785
\bibitem[Igumenshchev et al.(1996)]{ig96} Igumenshchev I.V., Chen X., Abramowicz M.A.,  1996, \mnras, 278, 2361
\bibitem[Igumenshchev et al.(1998)]{ig98} Igumenshchev I.V., Abramowicz M.A., Novikov I. D., 1998, \mnras, 298, 1069
\bibitem[Kato and Nakamura(1998)]{ka98} Kato S., Nakamura K.E., 1998, \pasj, 50, 559
\bibitem [Lanzafame et al.(1998)]{la98} Lanzafame G., Molteni D.,  Chakrabarti, S. K., 1998,  \mnras, 299, 799
\bibitem[Liang and Thompson(1980)]{li80} Liang E.P.T., Thompson K.A.,  1980, \apj , 240, 271
\bibitem [Lu et al.(1999)]{lu99} Lu Ju-Fu, Gu Wei-Min, Yuan Feng, 1999, \apj,  523, 340
\bibitem[Nakamura et al.(1996)]{na96} Nakamura K.E., Matsumoto R., Kusunose M., Kato S., 1996, \pasj, 48, 761
\bibitem[Narayan and Yi(1995)]{narai95} Narayan R., Yi I., 1995, \apj, 444, 231
\bibitem[Narayan and Mahadevan(1995)]{naram95} Narayan R., Mahadevan R., 
1995, Nature, 374, 623
\bibitem[Narayan et al. (1997)]{naraa97} Narayan R., Kato S., and Honma F., 1997, \apj, 476, 49
\bibitem [ Paczy{\'{n}}ski and Wiita (1980)]{pa80}Paczy{\'{n}}ski B.,Wiita P.J., 1980, \ A\&A,  88, 23
\bibitem[Shakura and Sunyaev(1973)]{ss73} Shakura N.I., Sunyaev R.A., 1973, \ A\&A, 24, 337

\end{thebibliography}
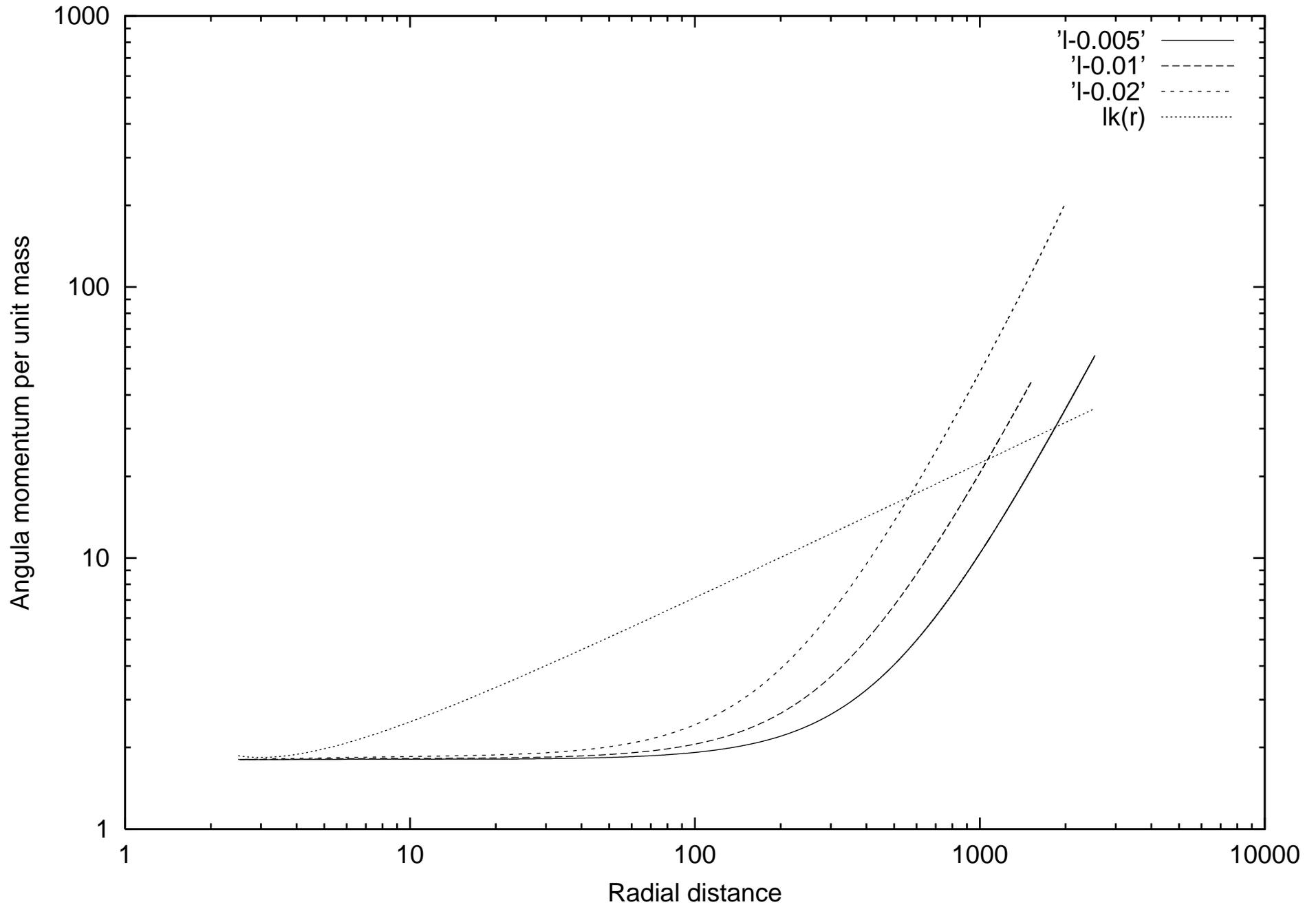
\end{document}